\documentclass{article}
\usepackage[utf8]{inputenc}
\usepackage{url}
\usepackage{graphicx,algorithm,algcompatible,amsmath}
\usepackage{hyperref}
\usepackage{subfigure}
\usepackage{algpseudocode}
\usepackage{amsthm,amssymb,inputenc}
\usepackage[T1]{fontenc}

\title{Analysis of Greenhouse Gases\\
\large Final Project\\
\large Stochastic Processes 625.722\\
\large Johns Hopkins University}
\author{Shalin Shah\\sshah100@jhu.edu}
\begin{document}

\maketitle
\begin{abstract}
Climate change is a result of a complex system of interactions of greenhouse gases (GHG), the ocean, land, ice, and clouds. Large climate change models use several computers and solve several equations to predict the future climate. The equations may include simple polynomials to partial differential equations. Because of the uptake mechanism of the land and ocean, greenhouse gas emissions can take a while to affect the climate. The IPCC has published reports on how greenhouse gas emissions may affect the average temperature of the troposphere and the predictions show that by the end of the century, we can expect a temperature increase from $0.8^{\circ}$ C to $5^{\circ}$ C. In this article, I use Linear Regression (LM), Quadratic Regression and Gaussian Process Regression (GPR) on monthly GHG data going back several years and try to predict the temperature anomalies based on extrapolation. The results are quite similar to the IPCC reports.
\end{abstract}
\emph{Keywords}: global warming, extrapolation, linear model, quadratic model, gaussian process regression, climate change, greenhouse gases, machine learning
\section{Introduction}
The climate is a result of complex interactions between several elements. Greenhouse gases and the sun are both equally responsible for maintaining a temperature at which we can live, in the troposphere. The sun continually emits UV and IR radiation, some of which is reflected back from the Ozone layer and also from the ice in the Arctic and the Antarctic. The clouds and the land also reflect sun rays. This reflection is called albedo which is the reason for the temperate climate that our planet has. Greenhouse gases like CO2, CH4 and water vapor absorb some of the sun heat and cause a warming of the atmosphere which then can support various species of plants and animals. However, mostly because of fossil fuel burning which emits CO2 and CH4, the amount of greenhouse gases in the atmosphere is increasing which is causing the temperature to gradually increase. The emitted CO2 and CH4 are also absorbed by land and the ocean, which is called uptake. But the absorbed greenhouse cases may also be then released back into the atmosphere, and so there is a gradual stabilization in the amount of greenhouse gases at a higher level than the present because of anthropogenic emissions \cite{maslin2014climate} \cite{maslin2013climate} \cite{maslin2008global}.\\\\
In this article, I attempt to use Linear Regression (LM), Quadratic Regression, and Gaussian Process Regression (GPR) \cite{rolnick2019tackling} \cite{williams2006gaussian} \cite{zheng2018analysis} to predict how the levels of GHG affect the average temperature of the atmosphere through temperature anomalies. Note that the effect of carbon emissions in one area of the world affects the entire world if sufficient time is given to the atmospheric and oceanic forces to stabilize.
\section{Greenhouse Gas Models and Emission Models}
Table 1 shows the correlation matrix and table 2 shows various models with the R-Squared on a test set. The correlation matrix shows a strong correlation between CO2 and the temperature anomalies as well as between CH4 and the temperature anomalies. CO2 is more abundant in the atmosphere and is a stronger indicator of temperature anomalies as compared to CH4. Humidity has a small correlation, but the other greenhouse gases have a stronger correlation.\\\\
I tried several models to see the effect of increasing greenhouse gas concentration in the atmosphere. The results section follows this section. I tried linear regression, non-linear quadratic regression and Gaussian process regression (GPR). All three models are able to extrapolate beyond what is in the training data.\\\\
The greenhouse gas models try to predict what would be the anomaly in temperature when the greenhouse gas concentration in the atmosphere is changed to a lower and higher multiple of the concentration on 10/2017. On 10/2017, the CO2 concentration (as measured by Mauna Loa Observatory) was 404 PPM, the CH4 concentration was 1858 PPB and Humidity was 65.4. The temperature anomaly on 10/2017 was 0.90. This means that compared to the expected temperature, this month was warmer by $0.90^{\circ}$ C. This threshold for extrapolation is used in the greenhouse gas models.\\\\
The CO2 level on 7/1991 was 356 PPM and the CH4 level was 1716 PPB, and the relative humidity level was 53.4. The temperature anomaly on 7/1991 was $0.47^{\circ}$ C. This threshold for extrapolation is used in the Emission models.\\\\
See figures 1 through 6.
\section{Results}
\subsection{The Data and Packages}
The temperature anomaly data is taken from:\\ \url{https://climate.nasa.gov/vital-signs/global-temperature/}.\\\\
The GHG data is taken from:\\
\url{https://www.esrl.noaa.gov/gmd/dv/}. (Mauna Loa Observatory)\\\\
The CO2 emissions data is taken from:\\
\url{https://datahub.io/core/co2-fossil-global}\\\\
Other data is available here:\\
\url{https://esgf-node.llnl.gov/search/cmip5/}.\\\\
We use the mlegp package for Gaussian Process Regression \cite{dancik2013mlegp}.\\\\
We use lm and nls in R \cite{RProject} for linear and non-linear regression:\\
\url{https://cran.r-project.org/}
\subsection{Scatter Plot of Greenhouse Gases}
Figure 1 shows a scatter plot of standardized greenhouse gases with the temperature anomaly. The trend shows that as the concentration of greenhouse gases increase, the temperature anomaly also increases. The effect is strongest with methane (CH4) but because CH4 is present in only low quantities in the atmosphere, the effect it has on temperature is not very strong. CO2 and humidity also have a positive slope.
\subsection{Extrapolation of Greenhouse Gases}
Figure 2 and Figure 3 are plots with extrapolation on the model. Starting from a multiplier of 0.02 and in increments of 0.001 I query the model with values of the greenhouse gases multiplied by the multiplier. Figure 2 shows the results of a linear model while Figure 3 shows the results in a non-linear regression model (quadratic).\\\\
Figure 2 shows that as the greenhouse gas levels increase, the temperature anomaly also increases. For instance, if the CO2 level is increased 1.5 times as compared to the 10/2017 level (400 PPM to 600 PPM), the temperature anomaly will be about $2.5^{\circ}$ C. The same increase in CH4 causes a temperature anomaly of $5^{\circ}$ C, but this is relatively less important than the CO2 levels (unless there are gas hydrate eruptions in the ocean). The plot also shows that if the CO2 level was decreased to half (200 PPM), the temperature anomaly will be $-1.25^{\circ}$ C. If the CO2 level was doubled to 800 PPM, the temperature anomaly will be $5^{\circ}$ C. This is easily reconciled with the IPCC reports which have almost the same results.\\\\
I also tried to fit a non-linear quadratic regression model as shown in figure 3. The results are similar. But as figure 3 shows, the quadratic has a much larger curvature in the initial stages and increases to almost linear after a multiplier of 1.5. This is clearly what the scientists expect in that there is a tipping point after which the temperature anomaly increases more rapidly.\\\\
Figures 7 and 8 show the extrapolation charts for Gaussian Process Regression (GPR). It shows that if the CO2 level is increased to 600 PPM, the temperature will increase by $3^{\circ}$ C. For CH4, GPR shows a slightly higher increase of $5^{\circ}$ C if the level is increased by $1.5$ times the level on 10/2017 i.e. increase the CH4 to 2787 PPB.
\subsection{Analysis of Emissions}
As we increase carbon emissions, the CO2 and CH4 levels in the atmosphere increase. The figures 4,5 and 6 show some analysis of emissions and the CO2 level as compared to the level on 7/1991 (356 PPM).\\\\
Figure 4 shows a linear fit and a scatter plot of CO2 levels and the emissions. Figure 5 shows the results of fitting a linear model with increasing emissions. It shows that if we increase the emissions 1.5 times that on 7/1991, we can expect the CO2 level in the atmosphere to increase to about 390 PPM (which we have already crossed). It also shows that if we decrease the emissions by half, the CO2 level will drop to about 330 PPM (not a very large decrease). If we reduce the emissions to $0$, the CO2 level will decrease to about 310 PPM (as compared to the 7/1991 levels).\\\\
Figure 5 shows the results of fitting a quadratic regression model, with extrapolation on the emissions. The results are quite similar. But as noted on the greenhouse gas models, the curvature is larger initially. But this model shows that if we reduce emissions to $0$, the CO2 level will stabilize to about 330 PPM, a slightly different result than the linear model.
\subsection{Discussion}
As noted in the previous section, the levels of greenhouse gases in the atmosphere is increasing at a rapid rate, thus causing a proportionate increase in the temperature. Global warming can cause many undesirable things like the following \cite{maslin2008global}:
\begin{enumerate}
    \item Increase in global temperatures by $0.8^{\circ} C$ to $5^{\circ} C$.
    \item Increase in the sea level, causing undesirable effects on coastal cities.
    \item Increase in the frequency of severe weather events like storms, droughts and heat waves and the severity of winters.
    \item Decrease in the world forests, which will cause a feedback effect.
    \item Release of CH4 from deep ocean gas hydrate deposits.
    \item Increase in epidemics of vector borne diseases like malaria because of a proliferation of disease spreading insects like mosquitoes.
    \item Negative effect on farming, thus causing food shortage
    \item Further shortage in the availability of clean drinking water
\end{enumerate}
\begin{figure}[ht]
\label{cc_6}
\centering
\includegraphics[scale=0.9]{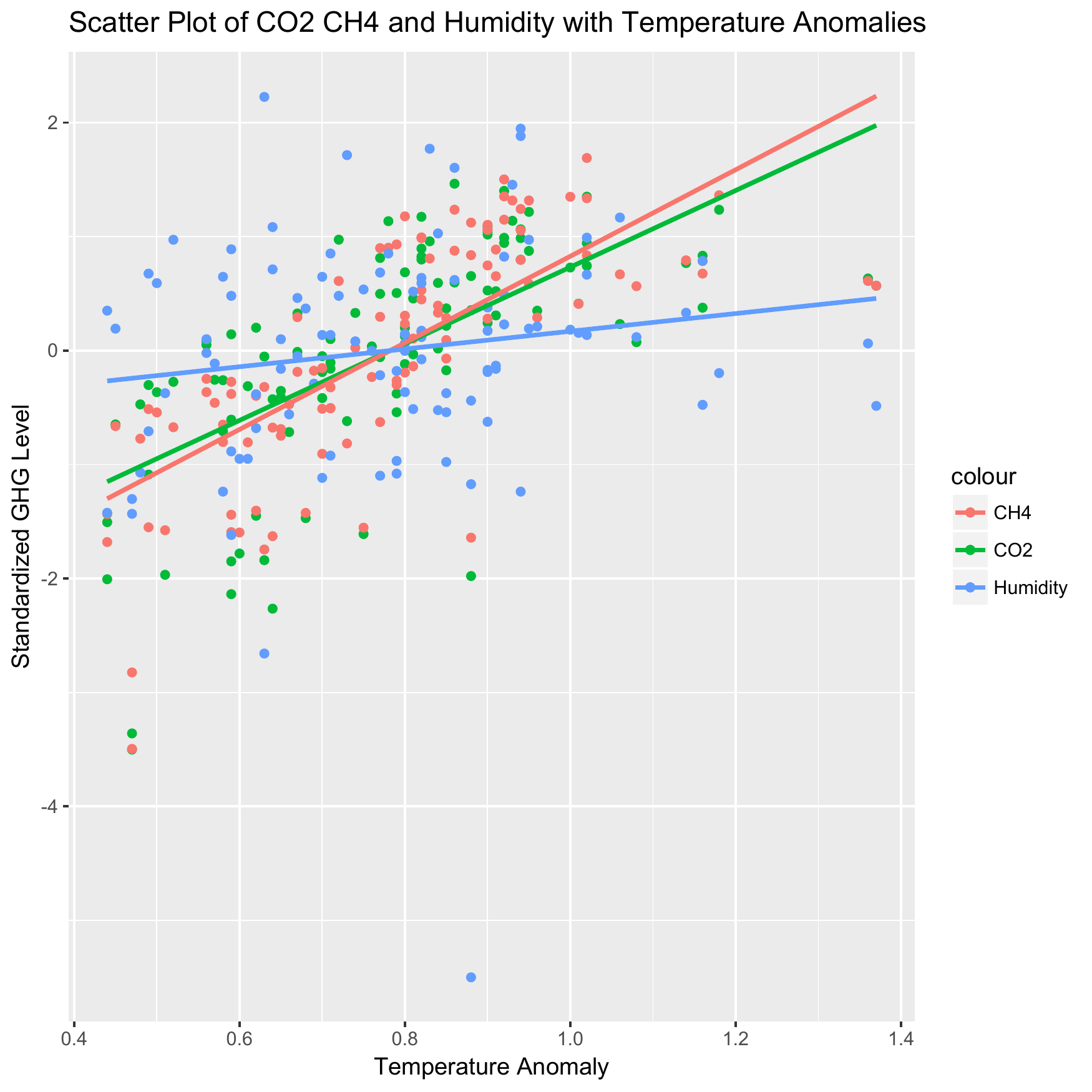}
\caption{Scatter Plot of Greenhouse Gases and Temperature Anomalies}
\end{figure}

\begin{figure}[ht]
\centering
\includegraphics[scale=0.9]{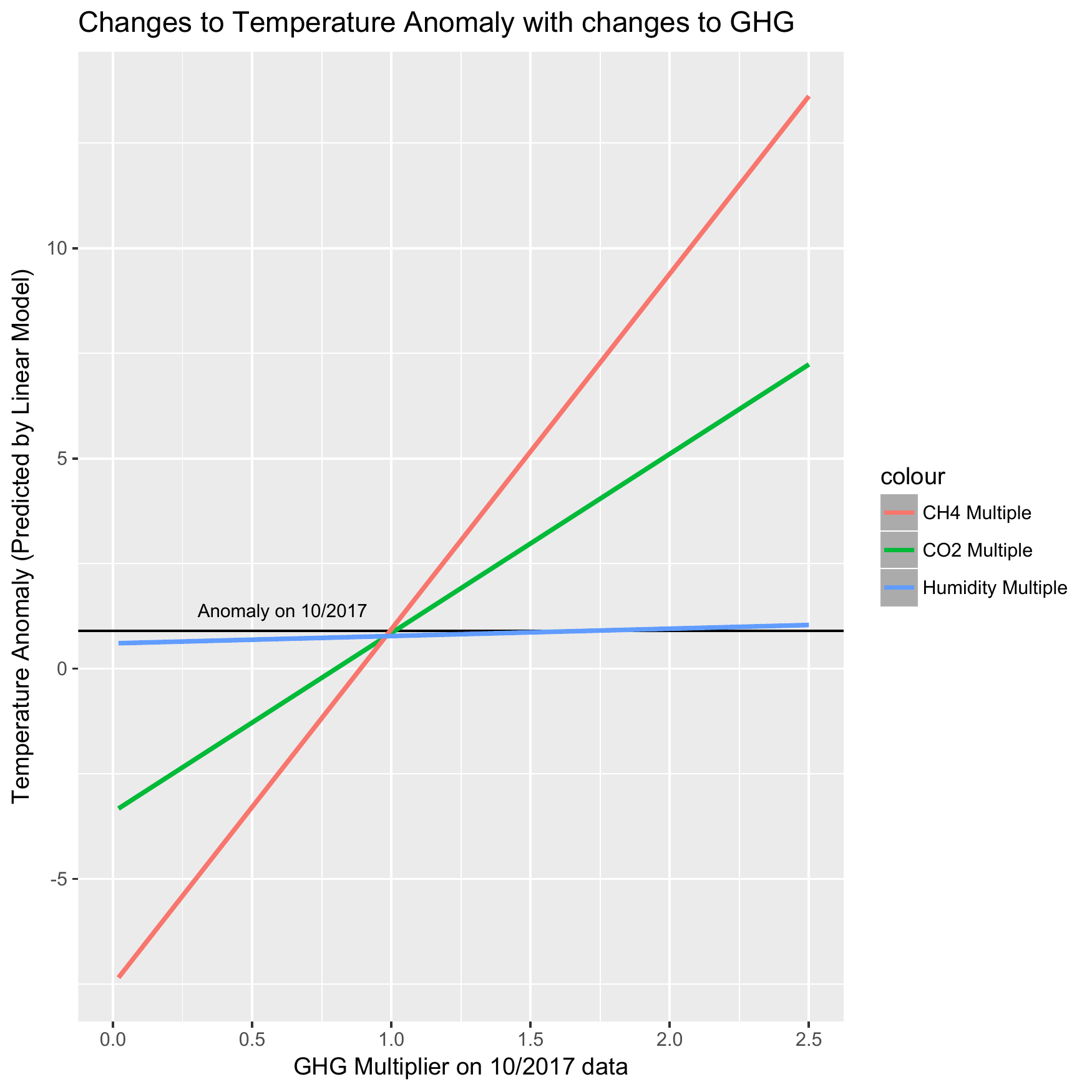}
\caption{Extrapolation by Scaling GHG Levels using LM}
\end{figure}

\begin{figure}[ht]
\centering
\includegraphics[scale=0.9]{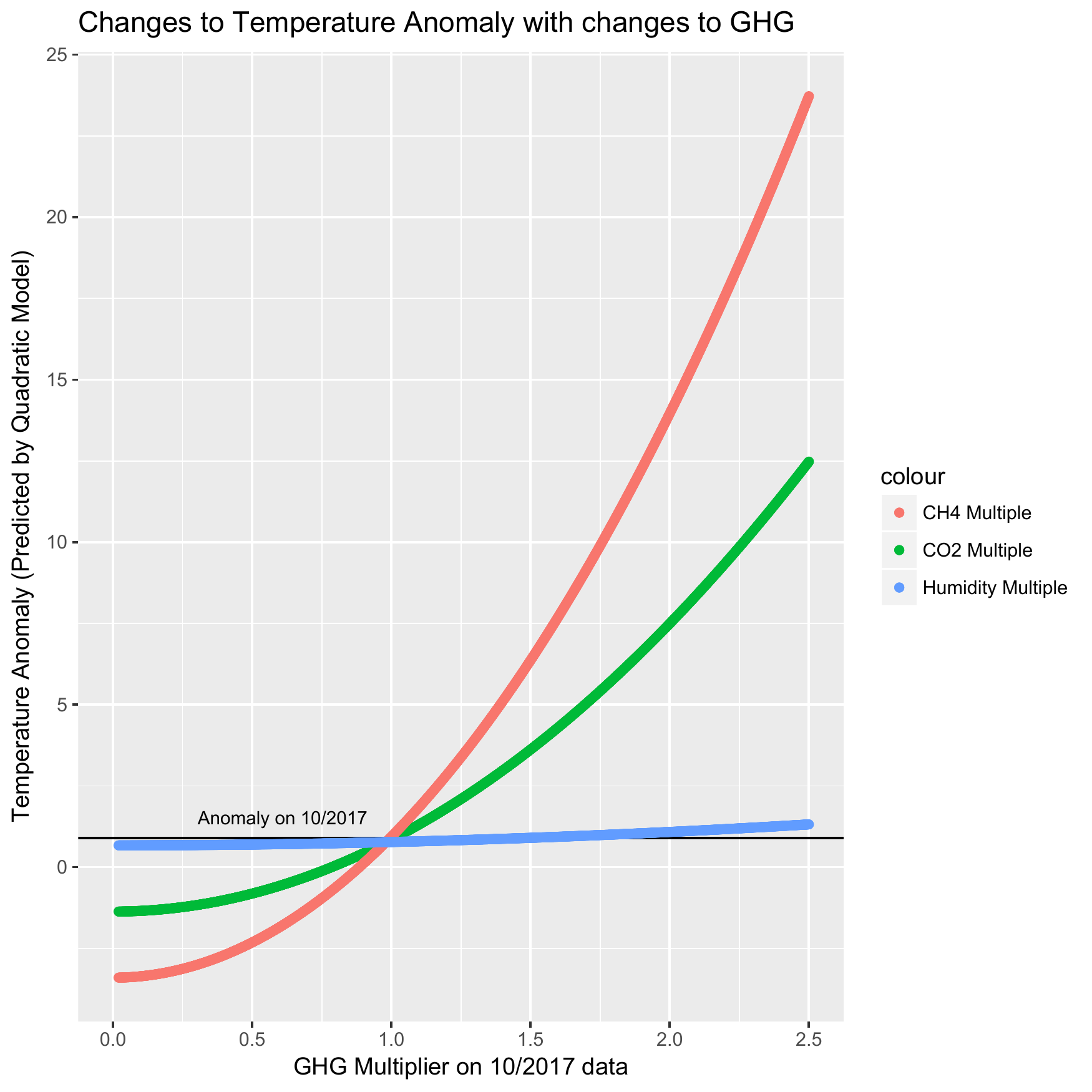}
\caption{Extrapolation by Scaling GHG Levels using a Quadratic Model}
\end{figure}

\begin{figure}[ht]
\centering
\includegraphics[scale=0.9]{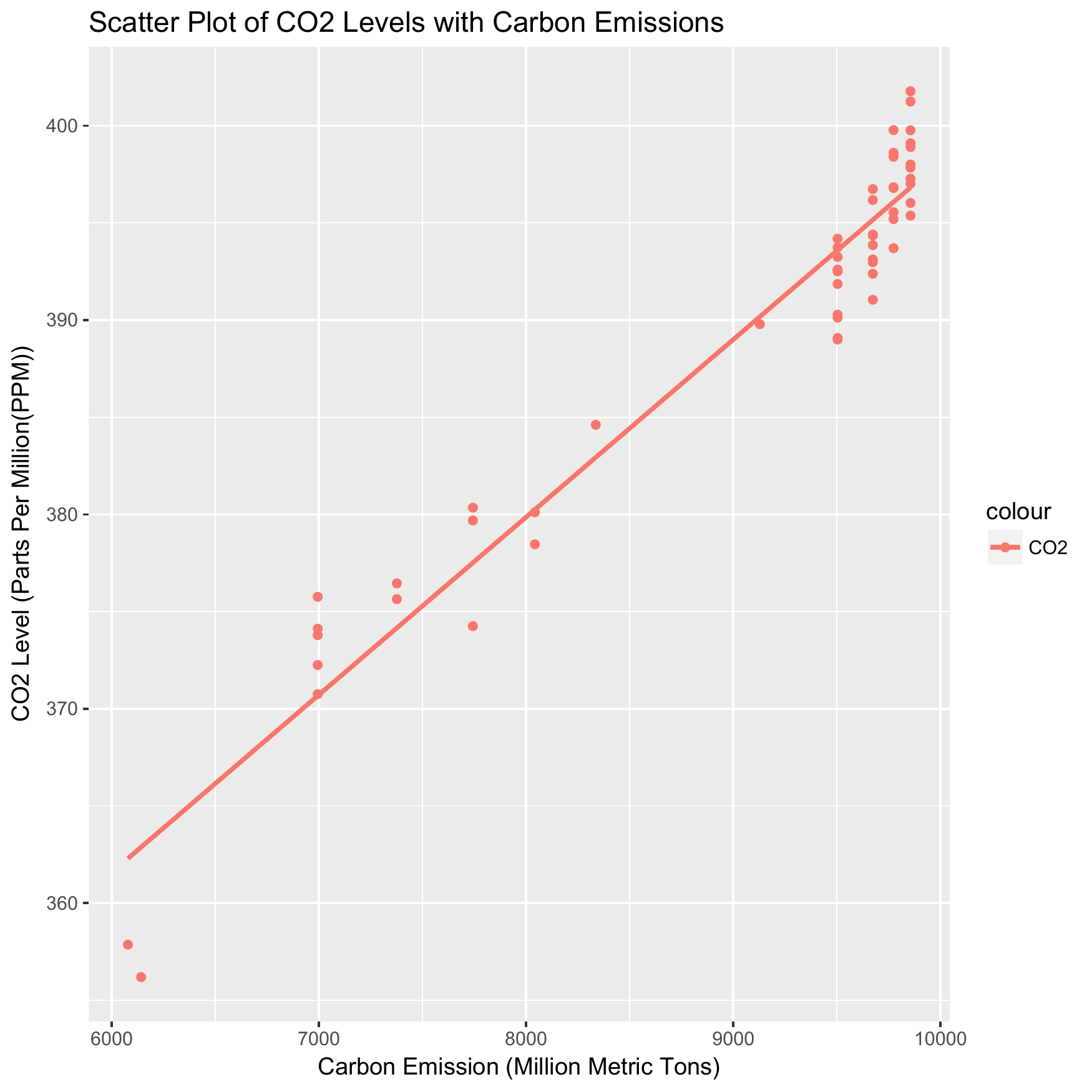}
\caption{Scatter Plot of Emissions and CO2 Levels}
\end{figure}

\begin{figure}[ht]
\centering
\includegraphics[scale=0.9]{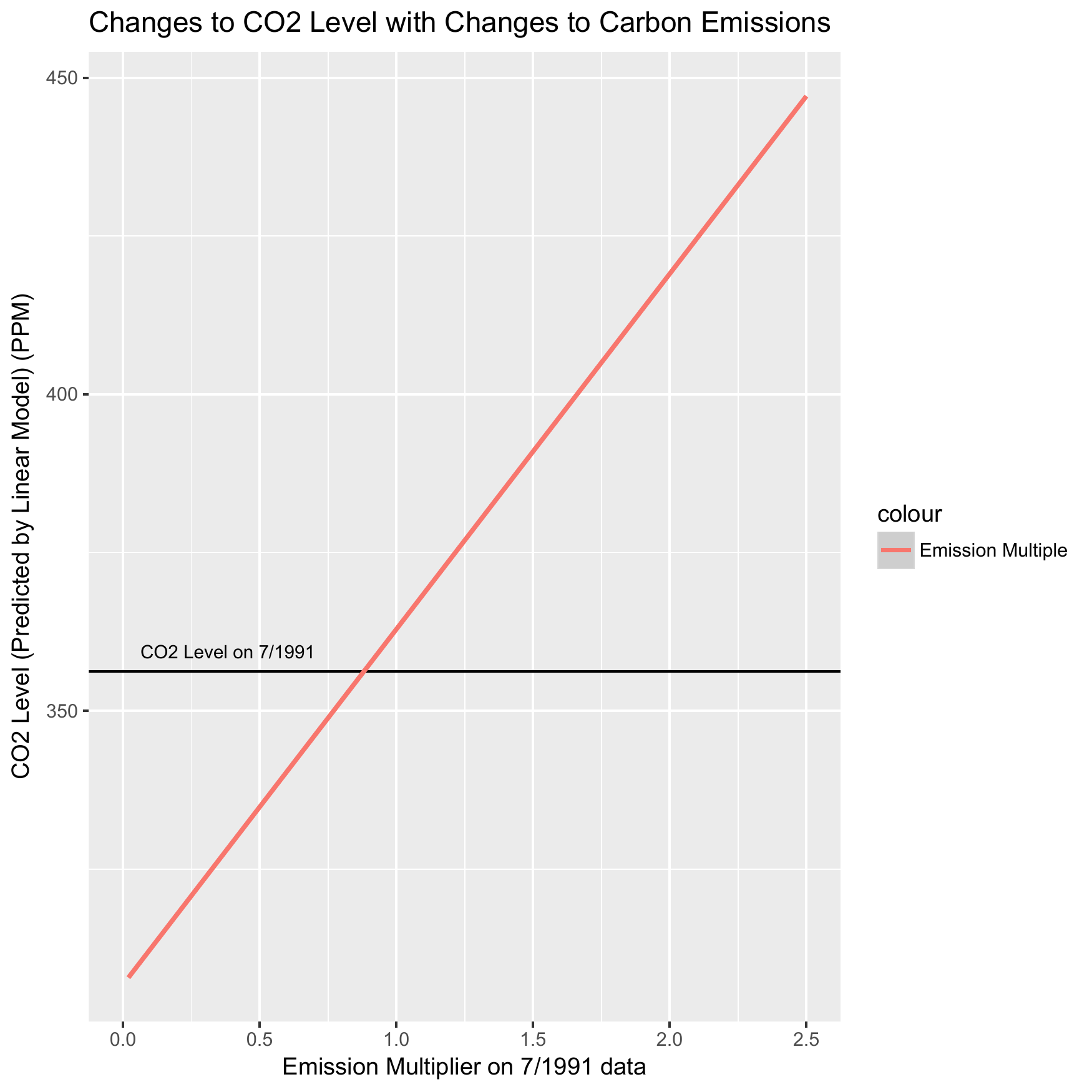}
\caption{Extrapolation by Scaling Emissions using LM}
\end{figure}

\begin{figure}[ht]
\centering
\includegraphics[scale=0.9]{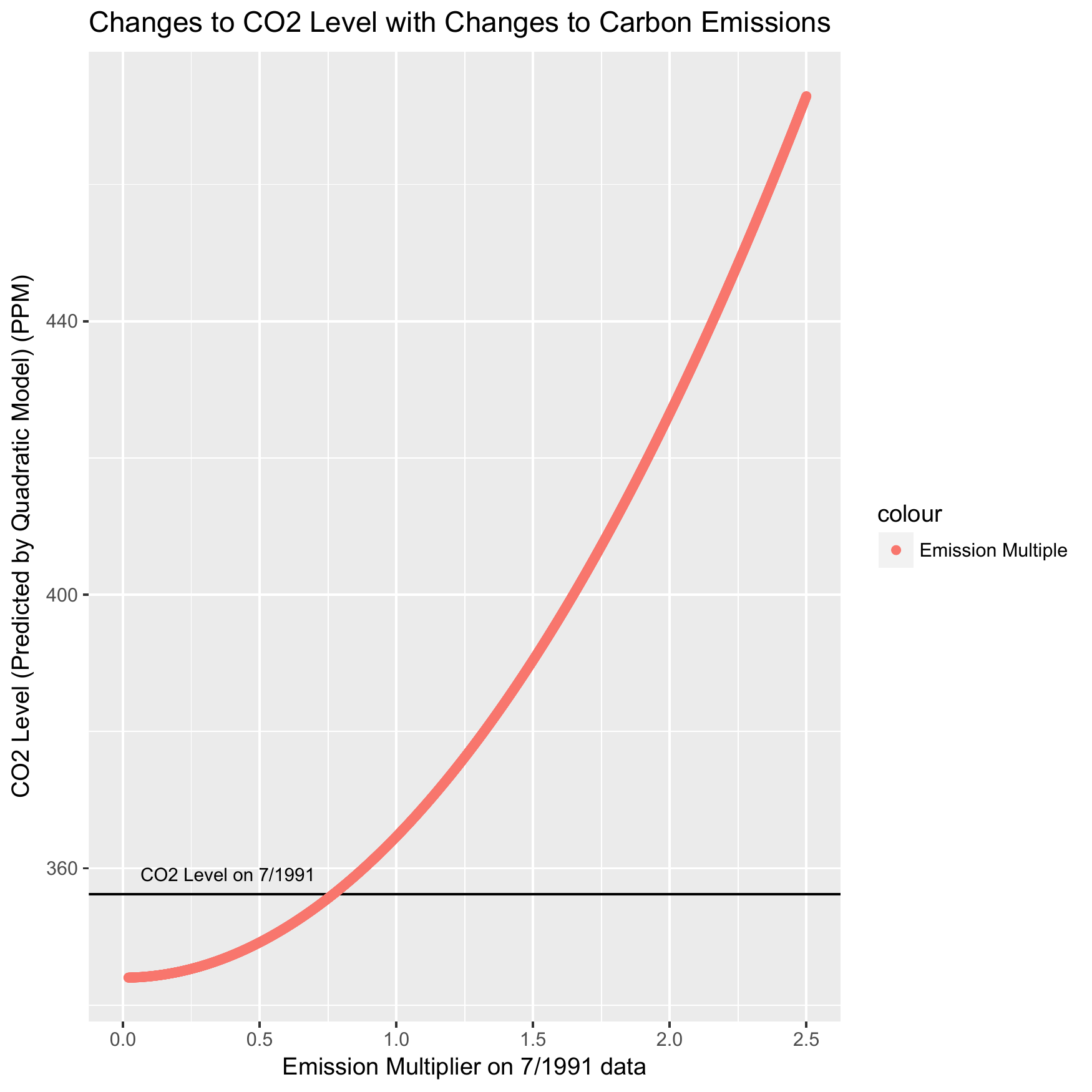}
\caption{Extrapolation by Scaling Emissions using a Quadratic Model}
\end{figure}

\begin{figure}[ht]
\centering
\includegraphics[scale=0.9]{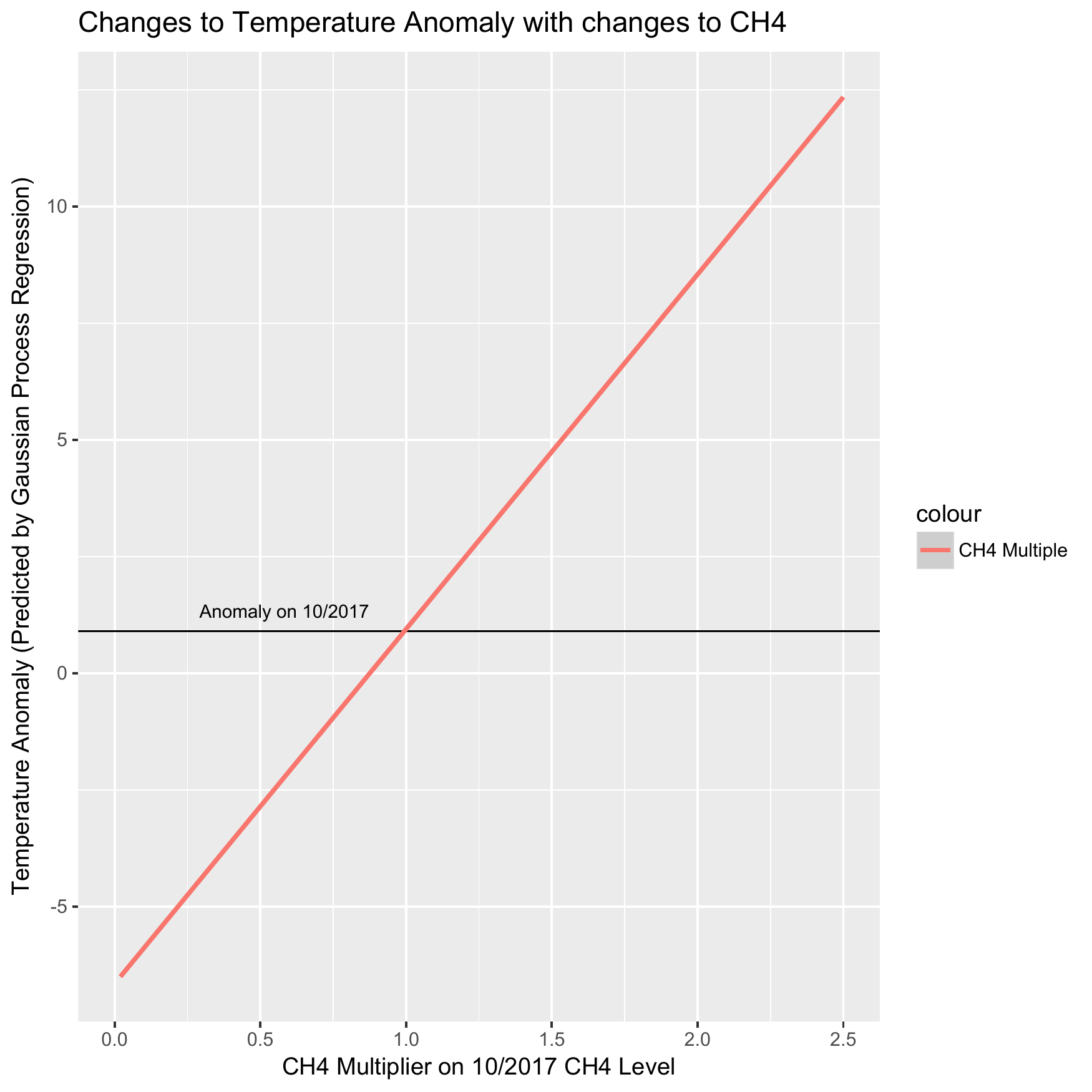}
\caption{Extrapolation by Scaling the CH4 Level using Gaussian Process Regression}
\end{figure}

\begin{figure}[ht]
\centering
\includegraphics[scale=0.9]{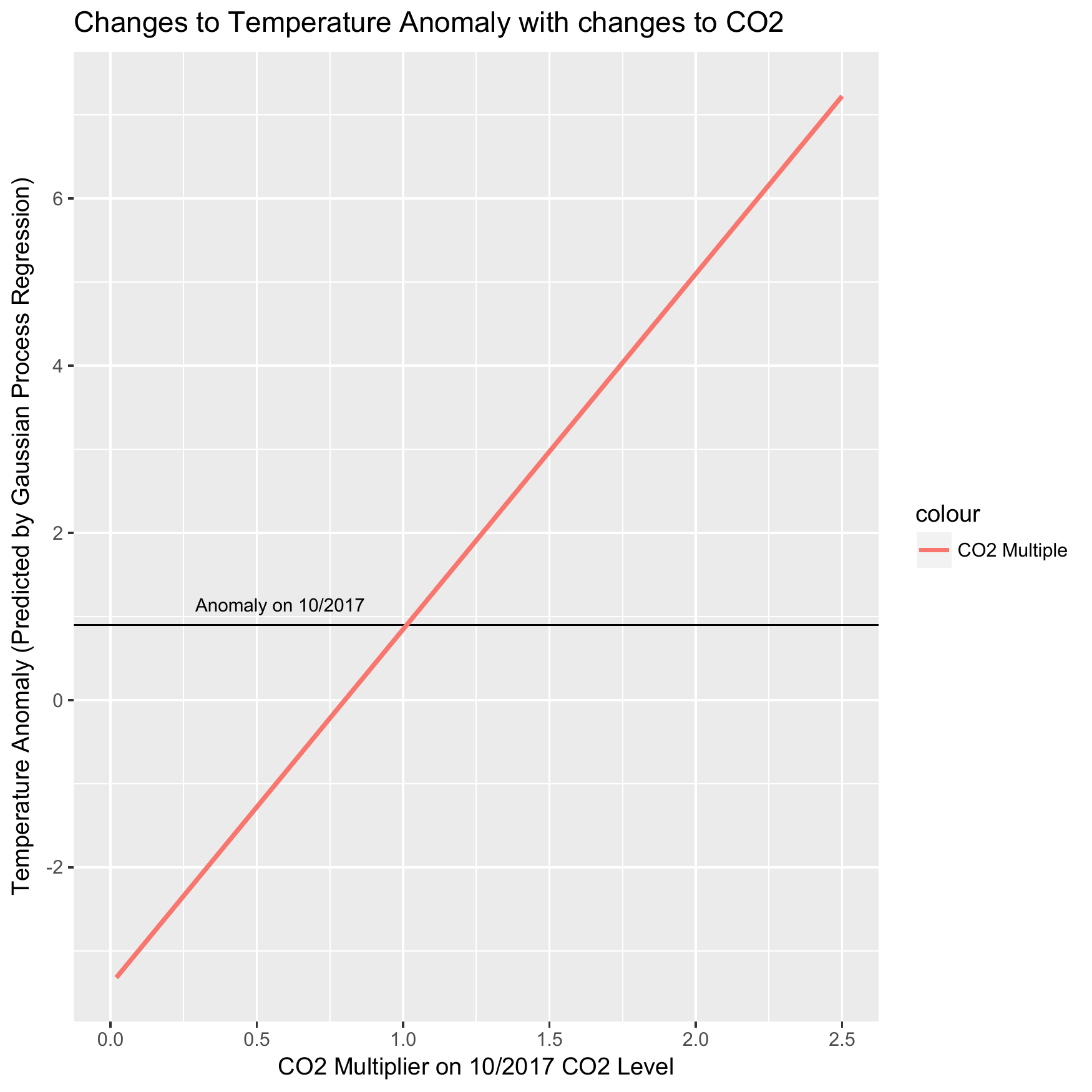}
\caption{Extrapolation by Scaling the CO2 Level using Gaussian Process Regression}
\end{figure}

\begin{table}[t]
\caption{Correlation Matrix}
\begin{tabular}{p{2cm}p{2cm}p{2cm}p{2cm}p{2cm}}
\hline
Greenhouse Gases &CO2 &CH4 &Humidity &Temperature Anomaly\\
\hline
CO2 &1 &0.94 &0.42 &0.65\\
CH4 &0.94 &1 &0.38 &0.73\\
Humidity &0.42 &0.38 &1 &0.15\\
Temperature Anomaly &0.65 &0.73 &0.15 &1\\
\end{tabular}
\end{table}

\begin{table}[t]
\caption{Temperature Anomalies and GHG Levels}
\begin{tabular}{p{4cm}p{4cm}p{4cm}}
\hline
Date &Name &Value\\
\hline
7/1991 &CO2 &356 PPM\\
7/1991 &CH4 &1716 PPB\\
7/1991 &Humidity &53.4\\
7/1991 &Temperature Anomaly &$0.47^{\circ}$ C\\
\hline
10/2017 &CO2 &404 PPM\\
10/2017 &CH4 &1858 PPB\\
10/2017 &Humidity &65.4\\
10/2017 &Temperature Anomaly &$0.90^{\circ}$ C
\end{tabular}
\end{table}

\begin{table}[t]
\caption{Model Predictions}
\begin{tabular}{p{1cm}p{2cm}p{2cm}p{3cm}p{3cm}}
\hline
GHG &Model &Multiplier &GHG Level Change &Temperature Anomaly\\
\hline
CO2 &Linear &2 &404 to 808 PPM &$5^{\circ} C$\\
CO2 &Quadratic &2 &404 to 808 PPM &$7.5^{\circ} C$\\
CO2 &GPR &2 &404 to 808 PPM &$5^{\circ} C$\\
\hline
CH4 &Linear &2 &1858 to 3716 PPB &$9^{\circ} C$\\
CH4 &Quadratic &2 &1858 to 3716 PPB &$12.5^{\circ} C$\\
CH4 &GPR &2 &1858 to 3716 PPB &$8.5^{\circ} C$\\
\hline
CO2 &Linear &1.5 &404 to 606 PPM &$2.5^{\circ} C$\\
CO2 &Quadratic &1.5 &404 to 606 PPM &$3.75^{\circ} C$\\
CO2 &GPR &1.5 &404 to 606 PPM &$3^{\circ} C$\\
\hline
CH4 &Linear &1.5 &1858 to 2787 PPB &$5^{\circ} C$\\
CH4 &Quadratic &1.5 &1858 to 2787 PPB &$6.25^{\circ} C$\\
CH4 &GPR &1.5 &1858 to 2787 PPB &$5^{\circ} C$\\
\hline
CO2 &Linear &0.5 &404 to 202 PPM &$-1.25^{\circ} C$\\
CO2 &Quadratic &0.5 &404 to 202 PPM &$-1^{\circ} C$\\
CO2 &GPR &0.5 &404 to 202 PPM &$-1^{\circ} C$\\
\hline
CH4 &Linear &0.5 &1858 to 929 PPB &$-3.75^{\circ} C$\\
CH4 &Quadratic &0.5 &1858 to 929 PPB &$-2.5^{\circ} C$\\
CH4 &GPR &0.5 &1858 to 929 PPB &$-2.5^{\circ} C$\\
\end{tabular}
\end{table}

\begin{center}
\begin{table}[t]
\caption{R-Squared of Various Models}
\begin{tabular}{p{6cm}p{6cm}}
\hline
{\bf Model} &{\bf R-Squared}\\
\hline
CO2 Linear Model &0.72\\
CH4 Linear Model &0.83\\
Humidity Linear Model &0.23\\
Combined Linear Model &0.84\\
CO2 Quadratic Model &0.73\\
CH4 Quadratic Model &0.83\\
Humidity Quadratic Model &0.20\\
Combined Quadratic Model &0.84\\
GPR CO2 Model &0.82\\
GPR CH4 Model &0.81\\
GPR Humidity Model &-0.31\\
GPR Combined Model &0.73\\
\end{tabular}
\end{table}
\end{center}

\section{Conclusion}
In this article, I analyzed some data of the greenhouse gases CO2, CH4 and Humidity and I find that all three correlate well with NASA's temperature anomaly data set. Correlation is not causation, but it has been noted through experiments in laboratories that greenhouse gases absorb the heat of the sun and so increase the atmospheric temperature. It is also conjectured that one of the reasons for the ice ages in the past is the reduction in the greenhouse gases \cite{maslin2008global}.\\\\
Through extrapolation on the models, I was able to predict what the temperature anomaly will be across levels of CO2 and CH4 multiplied by a multiplier. I note the results in the results section of this article. \\\\
I found it difficult to collect and process data from disparate sources because of availability and unknown formats, which is why I was not able to use other variables like cloud, land and ice albedo, ocean indicators etc. which are crucial to understanding global warming. Future work will include these variables and create more accurate predictions of climate change.
\nocite{*}
\bibliographystyle{unsrt}
\bibliography{GreenhouseGases}
\section{Appendix}
\subsection{Appendix A: Linear Regression}
A linear model is linear in the parameters and variables. It models the response variable in the following form:\\\\
$y=x_i\beta + \epsilon$\\\\
Where $\epsilon$ is $0$ mean Gaussian noise.\\\\
In a likelihood formulation, maximizing the posterior is equivalent to maximizing the likelihood. In a linear regression model, the likelihood is Gaussian.\\\\
$P(Model|Data) \sim P(Data|Model) = \prod_i N(\mu_i, \sigma^2)$\\\\
Where, $\mu_i=x_i\beta$ and $\sigma^2$ is a parameter which can also be parametrized. This is called a link function and in the case of linear regression, it is the identity link function. In the case of logistic regression, the link function is a sigmoid.\\\\
A prior $P(Model)$ can be placed in a Bayesian model (which I don't use in this article).
\subsection{Appendix B: Gaussian Process Regression}
Gaussian Process Regression (GPR) and kernel methods are similar. A kernel maps the input feature space into a possibly infinite dimensional feature space using basis functions. For instance, a quadratic regression model maps each dimension of the input into a polynomial equation of degree 2 (in our implementation, we omit the term with degree 1). Support vector machines are also an example of using the kernel trick.\\\\
GPR uses kernel methods with a Gaussian prior on the weights of the model. The kernel, or covariance function could be a linear kernel or maybe a squared exponential kernel as shown below:\\\\
$cov(f(x_p), f(x_q))=k(x_p, x_q)=exp\{-\frac{1}{2}|x_p-x_q|^2\}$\\\\
The covariance function represents a distribution over basis functions.\\\\
$f_*=N(\bf{0}$,$K(X_*, X_*))$\\\\
Where $f_*$ is a draw in the function space.\\\\
This is equivalent to Bayesian regression with an infinite dimensional feature space composed of basis functions of the original feature space. A thing to note that as in the squared exponential kernel above, kernel functions are are actually good similarity functions. Please see \cite{williams2006gaussian} for a highly detailed exposition of Gaussian processes for machine learning.
\end{document}